\documentclass[11pt,twoside]{article}
   \usepackage{newpasp}
   \usepackage{graphics}  % [pdftex]
   \usepackage{graphicx}
\index{anomalous X-ray pulsars}
\index{CGRO}
\index{gamma-rays}
\index{GLAST}
\index{neutron stars!magnetars}
\index{soft gamma-ray repeaters}

\font\fiverm=cmr5             \font\sevenrm=cmr7

\def\dover#1#2{\hbox{${{\displaystyle#1 \vphantom{(} }\over{
\displaystyle #2 \vphantom{(} }}$}}
{\catcode`\@=11                                                  
\gdef\SchlangeUnter#1#2{\lower2pt\vbox{\baselineskip 0pt\lineskip0pt    
\ialign{$\m@th#1\hfil##\hfil$\crcr#2\crcr\sim\crcr}}}}           
\def\gtrsim{\mathrel{\mathpalette\SchlangeUnter>}}               
\def\lesssim{\mathrel{\mathpalette\SchlangeUnter<}}    
\def\teq#1{$\, #1\,$}                         % text equation
\def\sigt{\hbox{$\sigma_{\hbox{\fiverm T}}$}}

\def\fsc{\alpha_{\hbox{\sevenrm f}}}                           
\def\erg{\varepsilon}
  
\def\gammax{\gamma_{\hbox{\sevenrm max}}}   
\def\gammadotC{\dot{\gamma}_{\hbox{\sevenrm C}}}
\def\Edot{\dot{E}_{\hbox{\sevenrm SD}}}

\def\edcomment#1{\iffalse\marginpar{\raggedright\sl#1\/}\else\relax\fi}
\reversemarginpar

\begin{document}
\title{Quiescent Magnetar Emission: Resonant Compton Upscattering}
\author{Matthew G. Baring}
\affil{Rice University, 
Department of Physics and Astronomy, MS-108,
P. O. Box 1892,
Houston, TX 77251-1892, USA\ 
\ \ {\rm [baring@rice.edu]}}

\begin{abstract}
A principal candidate for quiescent non-thermal gamma-ray emission from
magnetars is resonant inverse Compton scattering in the strong fields of
their magnetospheres.  This paper outlines expectations for such
emission, formed from non-thermal electrons accelerated in a pulsar-like
polar cap potential upscattering thermal X-rays from the hot stellar
surface.  The resultant spectra are found to be strikingly flat, with
fluxes and strong pulsation that could be detectable by GLAST.
\end{abstract}

\section{Introduction}

The quiescent, non-thermal X-ray emission of soft gamma repeaters (SGRs)
and anomalous X-ray pulsars (AXPs) is observed at levels that are very
intense for neutron stars.   It is quite possible that SGRs and AXPs
emit in a broad-band, pulsar-like mode, even if the power is not of a
rotational origin.  A principal candidate for such emission is inverse
Compton scattering, tapping abundant surface X-rays (inferred surface
temperatures for AXPs and SGRs are somewhat greater than for canonical
pulsars; see Perna et al. 2001) as targets for energetic electrons
accelerated in either polar cap or outer gap potentials.  Recently,
Cheng \& Zhang (2001) proposed a model for significant gamma-ray
emission from such magnetars, in the context of  the outer gap scenario.

Here, the possibility of magnetar gamma-ray emission from polar cap
models is explored.  In the strong fields of their magnetospheres, the
scattering is enhanced by the cyclotron resonance, thereby rendering
polar cap models even more promising for generating hard gamma-rays that
are potentially detectable by the upcoming Gamma-Ray Large Area Space
Telescope (GLAST) mission.   The present scenario is similar to previous
pulsar (e.g. Sturner, Dermer \& Michel 1995; Daugherty \& Harding 1996)
and old gamma-ray burst (Dermer 1990; Baring 1994) models. Primary
electrons with Lorentz factors up to some maximum \teq{\gammax} cool in
collisions with the X-ray photons, in the process generating a broad
spectrum extending into the gamma-ray band.

This paper outlines expectations for such emission spectra, specifically
incorporating relativistic cross sections at supercritical fields that
are germane to resonant Compton scattering in the magnetar regime. This
investigation will focus on a simple one-zone, uniformly-magnetized
region as a preparation for real geometry down the line, where the
complexity of pair cascading will be explored.  It is found that a very
flat soft gamma-ray continuum results, with spectral cutoffs that depend
strongly on the observational perspective to the field.  Spectra differ
significantly from those produced in outer gap models.  Estimates of the
anticipated flux indicate that GLAST may well detect magnetars in the
100 MeV -- 1 GeV band if they emit efficiently like gamma-ray pulsars.

\section{Resonant Compton Upscattering}

In strong fields the cross section for Compton scattering is resonant at
the cyclotron energy and a series of higher harmonics (e.g. see
Daugherty \& Harding 1986), effectively increasing the magnitude of the
process over the Thomson cross section \teq{\sigt} by of the order of
\teq{1/\fsc B}, where \teq{\fsc} is the fine structure constant.   Here,
as throughout the paper, magnetic fields are written in units of
\teq{B_{\rm cr}=m_e^2c^3/(e\hbar )=4.413\times 10^{13}}Gauss, the
quantum critical field strength.  In the non-relativistic, Thomson
regime (e.g. see Herold 1979), only the fundamental resonance is
retained.

The dominance by the resonance leads to an effective kinematic coupling
between the energies \teq{\erg_{\gamma}m_ec^2}  and \teq{\gamma m_ec^2}
of colliding photons and electrons, respectively, and the angle of the
initial photon \teq{\theta_{\gamma}} to the magnetic field lines: the
cyclotron fundamental is sampled when \teq{\gamma\erg_{\gamma}
(1-\cos\theta_{\gamma})=B}.  The simplicity of this coupling
automatically implies that integration over an angular distribution of
incoming photons results in a flat-topped emission spectrum for Compton
upscattering in strong magnetic fields.  This characteristic is
well-documented in the literature (e.g. see Dermer 1990; Baring 1994;
for old gamma-ray burst scenarios, and Sturner, Dermer \& Michel 1995
for pulsar contexts), specifically for collisions between
ultrarelativistic electrons and thermal X-rays emanating from a neutron
star surface.

For supercritical magnetar strength fields, the Thomson regime is no
longer operable, with relativistic corrections to the Compton cross
section becoming requisite.  These have been explored by Gonthier et al.
(2000) for cases appropriate to pulsars, namely when photons move
parallel to field lines in the electron rest frame (abbreviated ERF
hereafter).   The cross section is suppressed roughly by a Klein-Nishina
reduction when \teq{B\gtrsim 1} so that at \teq{B\sim 100} it becomes
comparable to \teq{\sigt} in the cyclotron resonance, and much smaller
for other energies.  The analytic developments of Gonthier et al. (2000)
for the resonant Compton cross section are employed in the computations
of this paper.

The development of emission spectra requires the determination of
equilbrium electron/pair distributions.  The resonant upscattering of
surface X-rays rapidly cools the electrons, resulting in the cessation
of acceleration in the polar cap potential gap at \teq{\gamma\lesssim
\gammax\sim 3\times 10^5}--\teq{10^6} (e.g. Sturner, 1995; Daugherty \&
Harding 1996; Harding \& Muslimov, 1998).  As the primary electrons
escape the gap, they continue to cool until the photon density is
diluted at altitudes above one stellar radius from the surface, when the
scattering kinematics also no longer sample the resonance.  The simplest
description of such cooling is explored here through standard kinetic
equations for development of the electron distribution.  The evolution
with altitude yields a mean electron distribution (i.e. averaged over
the electron path to higher altitudes) that is approximately
proportional to the inverse of the cooling rate \teq{\gammadotC}. This
amounts to a steep distribution (roughly \teq{\propto\gamma^{-4}}) for a
range in \teq{\gamma} of generally at least two decades below
\teq{\gammax}, with a very flat distribution (almost independent of
\teq{\gamma}) at lower energies. There is also a ``pile-up'' of
electrons at mildly-relativistic energies due to the concomitant
inefficiency of cooling. Such shapes can be directly inferred from
cooling profiles exhibited by Baring (1994) and Sturner (1995), who
specialized to the magnetic Thomson limit, and extend at least in
general nature to the relativistic domain, where the {\it slopes} of the
primary electron distribution are modified only slightly.

\vskip -0.8truecm
\resizebox{13.0truecm}{11.5truecm}{
\includegraphics{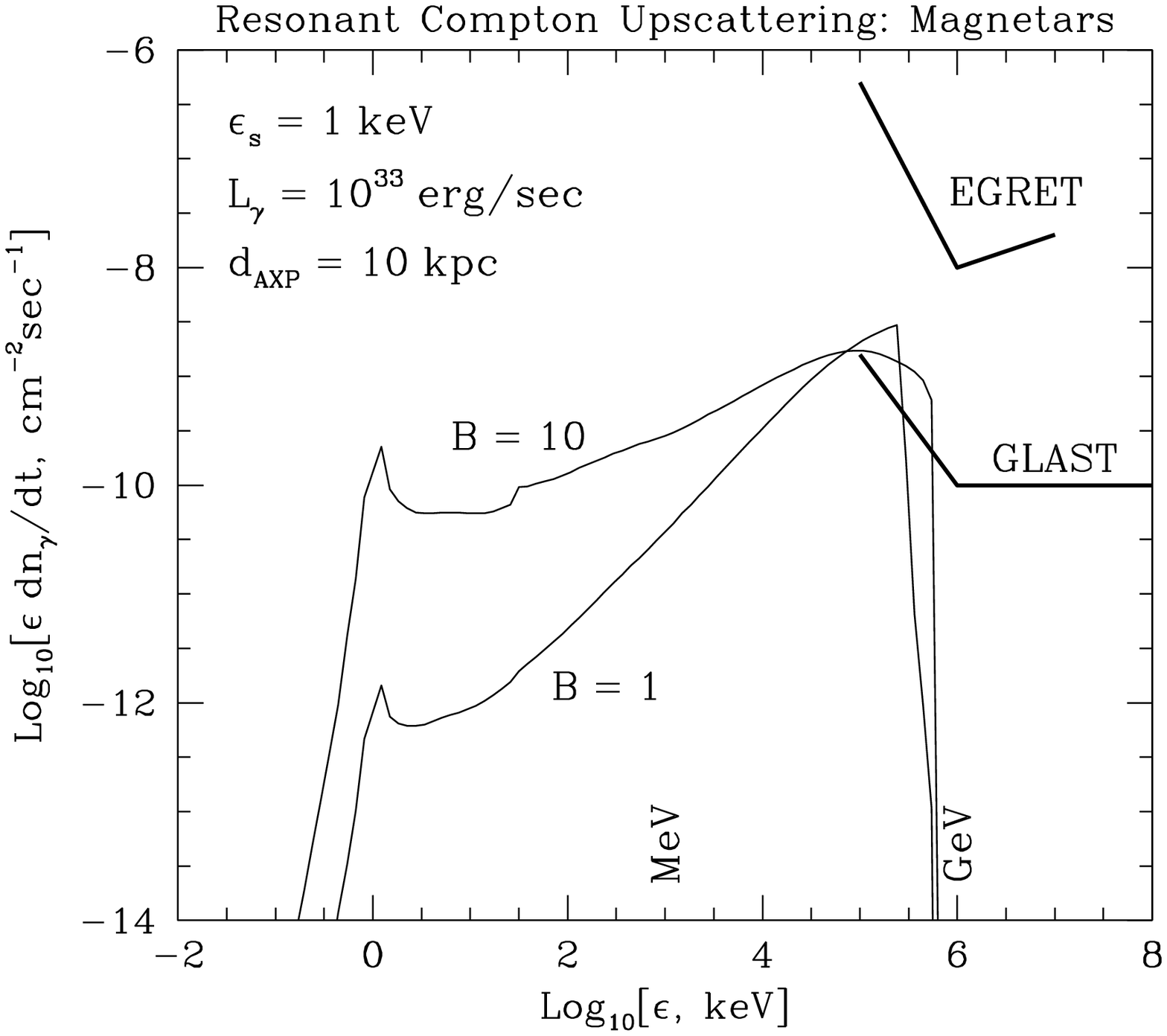}}

\begin{figure}[h]
\vspace{-2.3truecm}
\centerline{}
\caption{The angle-integrated \teq{F_{\nu}} spectra for resonant Compton
upscattering, for the two cases of \teq{B=1} and \teq{B=10}, with field
strengths \teq{B} expressed in terms of \teq{B_{\rm cr}=m_e^2c^3/(e\hbar
)=4.413\times 10^{13}}Gauss.  The spectra are normalized for a
luminosity of \teq{10^{33}}erg/sec (see text), and source distance of 10
kpc, typical for AXPs and SGRs.  An X-ray seed energy of
\teq{\erg_s=1}keV was used in the models.  The EGRET and GLAST flux
sensitivities are also depicted.
\label{fig:nuFnu_spec}
}
\end{figure}

This Compton-cooled distribution was then imported into standard
integrations for the resonant inverse Compton scattering spectrum, which
in the Thomson limit were detailed, for example, in Dermer (1990) and
Baring (1994). The principal complication incurred by introducing the
relativistic cross section is an extra integration as a consequence of
the finite energy change of photons in the Klein-Nishina domain; this is
handled routinely in the numerical integration routines.  Output results
of these integrations are exhibited in Fig.~\ref{fig:nuFnu_spec} for two
different field strengths. The \teq{B=1} case is similar in nature to
the Thomson regime results of Dermer (1990) and Baring (1994),
corresponding to a flat \teq{\erg^{-0}} distribution.  Klein-Nishina
effects manifest themselves in the \teq{B=10} case, that most
appropriate to magnetars, as a spectral steepening to around an
\teq{\erg^{-2/3}} spectrum.  Both cases extend out to photon energies
around 1 GeV, which reflect the upscattered energy from copious cooling
off surface X-rays by electrons with \teq{\gamma\sim\gammax}, These
photons are highly collimated almost along the magnetic field. Some pair
creation would be expected near this GeV cutoff (see the discussion of
Gonthier et al 2000); its exploration is deferred to future work.

The normalizations for the spectra in Fig.~\ref{fig:nuFnu_spec} are
established using an obvious observational benchmark.  Since a pulsar
mode for magnetar gamma-ray emission is considered here, the correlation
between spin-down power \teq{\Edot} and gamma-ray luminosity
\teq{L_{\gamma}^{\rm obs}} observed by the EGRET experiment on the
Compton Gamma-Ray Observatory (CGRO), namely
\begin{equation}
  L_{\gamma}^{\rm obs}\; \sim\; 1.7\times 10^{16}\;\sqrt{\Edot}
  \;\approx\; 5.3\times 10^{19}\,
       \dover{B_{\hbox{\sevenrm Gauss}}}{P^2_{\hbox{\sevenrm sec}}}
       \;\;\hbox{erg/sec}\quad ,
 \label{eq:spindown}
\end{equation}

can be {\it extrapolated} to the magnetar domain.  This equates to
roughly \teq{10^{32}}--\teq{10^{33}} erg/sec for AXPs and SGRs,
corresponding approximately to their \teq{\Edot}.  Note that this
luminosity is still two orders of magnitude smaller than the detected
quiescent X-ray luminosities, whose power cannot be of rotational
origin.

The spectra were normalized to this \teq{L_{\gamma}^{\rm obs}} roughly
around 1 GeV, for a source 10 kpc distant, resulting in predicted fluxes
that would be detectable by GLAST, a principal conclusion of this paper.
The flat spectra strongly contrast the outer gap predictions of Cheng \&
Zhang (2001), which are inherently steeper due to their use of
non-magnetic inverse Compton scattering.  The polar cap model therefore
indicates that EGRET would not have been expected to detect AXPs unless
the gamma-ray luminosity was comparable to the X-ray power.

Observe that potential GLAST detections would possess marked phase
dependence.  The kinematics of resonant scattering yields the strong
correlation \teq{\erg_{\gamma}\propto (1-\cos\theta_{\gamma} )^{-1}}
between the gamma-ray energy \teq{\erg_{\gamma}} and its angle to the
field \teq{\theta_{\gamma}}, translating into much softer spectra at
significant angles to the field, i.e. out of the main pulse. Another
discriminating feature is the expectation of strong polarization in the
gamma-ray signal, though this cannot be probed until future generation
missions.  It is anticipated that any GLAST detections of AXPs or SGRs
in quiescence will have a profound impact on our understanding of these
sources, with potential discrimination between outer gap and polar cap
models.

\end{document}